\documentclass{appolb}
\usepackage{amsmath}

\begin{document}
%% uncomment the following line to get equations numbered by (sec.num)
\eqsec

\title{On the effective actions for the spherical charged dust shell in General Relativity}

\author{Valentin D Gladush
\address{vgladush@gmail.com\\
Theoretical physics department, Dnepropetrovsk National University,\\
Gagarina Ave 72, Dnepropetrovsk, 49010, Ukraine}
\and
Alexander I Petrusenko
\address{iftifn@gmail.com\\
Theoretical physics department, Dnepropetrovsk National University,\\
Gagarina Ave 72, Dnepropetrovsk, 49010,
Ukraine}
}

\maketitle

\begin{abstract}
A simple and direct, based on the equations of motion, derivation of the variational principle and effective
actions for a spherical charged dust shell in general relativity is
offered.
This principle is based on the relativistic version of the
D'Alembert principle of virtual displacements and leads to the effective
actions for the shell, which describe the shell from the point of
view of the exterior or interior stationary observers. Herewith, sides
of the shell are considered independently, in the coordinates of the
interior or exterior region of the shell. Canonical variables for
a charged dust shell are built. It is shown that the conditions of
isometry of the sides of the shell lead to the Hamiltonian constraint
on these interior and exterior dynamical systems. Special cases of
the ``hollow'' and ``screening'' shells are briefly considered,
as well as a family of the concentric charged dust shells.
\end{abstract}


\section{Introduction}

The theory of spherically-symmetric thin shells plays key role for
construction of the effective non-trivial models for the collapsing gravitating
configurations. Thin shells have been found to have widespread application
in different areas of the General Relativity, astrophysics and cosmology
for modeling of the extended objects which thickness can be neglected.
For example, they are intensively used for the analysis of the basic
problems of a gravitational collapse, including its classical and
quantum aspects. In astrophysics, spherical shells are used for modeling
of the supernovas and other variable cosmic objects. At larger scales,
specific configurations of shells have also been considered to construct
cosmological models, to analyze phase transitions in the early universe
or to describe cosmological voids etc. (see for reviews \cite{ansoldi02,kijowski0}).

The equations of motion of spherical shells have been obtained in
\cite{israel}, the equations of motion of charged spherical shells
have been found in \cite{kuchar,deLaCruz,сhase,proszynski}. The construction
of the variational principle, as well as Lagrangian and canonical formalism for the shells and branes were discussed in \cite{farhi90}-\cite{gladGRG}.

Note that in most of the works mentioned, the variational principle
for the shell is constructed based on the general variational principle
containing the standard Einstein-Hilbert term, the actions of the substance
on the world sheet of the shell and suitable surface terms.
Subsequent reduction of the action of the shell leads to the action,
which  is based on gauge fixing and dependent
on the choice of the evolution parameter (generally proper time).

The approach to the construction of the variational principle
for spherical dust shells in terms of the stationary interior and
exterior observers was developed on the basic principles in \cite{gladJMa,gladGRG}.
This approach is based on a procedure for reduction of
the full action, which contains the Einstein-Hilbert terms for the
interior and exterior regions, the action for the dust matter on the
singular shell, the surface matching and normalizing terms, the surface
terms which are introduced to fix the metric on the boundary of the considered
region, and on the subsequent modification of the variational procedure.

In series of paper by J. Kijowski and collaborators \cite{kijowski2}-\cite{kijowski5}
the general approach to the construction of the Lagrangian and Hamiltonian variational principle for the composed ‘‘shell + gravity’’ system is proposed, starting
from first principles, without assuming any symmetry of the system.

In the case of a spherically symmetric shell in vacuum,
this formulation leads to a simple Hamiltonian system with
1 degree of freedom. The configuration variable is the area
of the shell, whereas the canonical momentum equals the
hyperbolic angle between the surfaces $t_{Schwarzschild}=const $ on one side and the surfaces $t_{Minkowski}=const $ on the other side of the shell. The Hamiltonian of the system is explicitly calculated in terms of the
“true degrees of freedom”, i.e. as a function on the reduced phase space.

Note that the approaches mentioned above are based on a cumbersome procedure for reduction of the full action, even in the case of spherical symmetry. It is even more complicated in presence of an electric charge. However, since generalized Birkhoff's theorem implies that spherically symmetric  gravitational field has no local degrees of freedom, one would expect that the system could be reduced to a single degree of freedom, the shell radius, with a potential obtained by solving equations for the gravitational and electromagnetic fields. Thus, at least for the region $r>r_{+}$, where $r_{+}$ is the external event horizon of the Reissner-Nordstr\"{o}m metric, we can the study dynamics of the charged shell in external or internal fixed gravitational and electromagnetic fields, i.e. on the Reissner-Nordstr\"{o}m geometries,  instead of the dynamics of the ‘‘charged shell + gravity+ electromagnetism’’ system. Fields inside and outside spherical shell are determined solely by the matching terms on the shell and the asymptotic behavior at infinity. There is no radiation, the full proper mass and charge of the dust shell are conserved, so no problem with the surface density of mass and charge and no need for detailed consideration of local values. In this case, one can use the known equations of motion of a charged spherical shell  \cite{kuchar,deLaCruz,сhase} to construct the effective actions. The construction of the effective actions for the spherical dust shell based on the equations of motion is ambiguous task and leads to different results depending on the choice of evolution parameter \cite{frolov}-\cite{berezin97}. In the majority of works the variational principle for shells
is constructed in the co-moving frame of reference, or in one of variants
of the freely falling frames of reference. In our opinion, the choice
of the exterior or interior remote stationary observer in the theory
of gravitating shells is the most natural and corresponds with the
real physics. The natural Hamiltonian formalism of a neutral self-gravitating
shell was considered in \cite{hajicek,dolgov}, where the Hamiltonian
of the shell is actually postulated.

In this paper we propose simple and direct construction of the variational
principle for charged dust spherical shells in general relativity.
This procedure is based on the generalization of the relativistic
version of the D'Alembert principle of virtual displacements \cite{Dalamb, Lanczos}.
It leads to two variational principles in the curvature coordinates for the internal and external regions of the shell.
As the result, the effective actions, Lagrangian and Hamiltonian for a charged spherical shell in the curvature coordinates of the interior and exterior regions of space-time are constructed, and the Hamiltonian constraint is obtained which plays the role of the integrals of motion.

Everywhere in this paper the gravitational constant $k=1$ and the
speed of light $c=1$. The metric tensor $g_{\mu\nu}\quad(\mu,\nu=0,1,2,3)$
has signature\\ (+ - - -).

\section{Spherically-symmetric space-time with a spherical shell}

Let us consider a spherically-symmetric compound configuration $D=D_{-}\cup\Sigma\cup D_{+}$
which is the union of concentric interior $D_{-}$ and exterior $D_{+}$
regions. These regions are matched together along time-like spatially
closed hypersurface $\Sigma$ which forms world sheet of a spherical
infinitely thin dust shell with dust density $\sigma$ and charge
density $\sigma_{e}$. Let $x^{i}:\{x^{2}=\theta,~x^{3}=\alpha\}\ (i,k=2,3)$
be the general angular, and $x_{\pm}^{a}(a,b=0,1)$ be the individual
coordinates defined in regions $D_{\pm}$. Then gravitational fields
in regions $D_{\pm}$ are generally described by metrics\begin{equation}
^{(4)}ds_{\pm}^{2}={}^{\left(4\right)}g_{\mu\nu}dx^{\mu}dx^{\nu}={}^{(2)}ds_{\pm}^{2}-r^{2}d\sigma^{2},\label{1.1}\end{equation}
\begin{equation}
^{(2)}ds_{\pm}^{2}=\gamma_{ab}^{\pm}dx_{\pm}^{a}dx_{\pm}^{b}\,,\qquad d\sigma^{2}=h_{ij}dx^{i}dx^{j}=d\theta^{2}+\sin^{2}\theta d\alpha^{2}.\label{1.2}\end{equation}
The two-dimensional metrics $\gamma_{ab}^{\pm}$ and the scale factor
$r$ are functions of the coordinates $x_{\pm}^{a}$. We have $^{(2)}ds_{+}|_{_{\Sigma}}={}^{(2)}ds_{+}|_{_{\Sigma}}={}^{(2)}ds_{+}$,
as well as the world line of the shell is set by the equation $x^{a}=x^{a}(s)$.
In the points on hypersurface we will define orthonormal basis \begin{equation}
\left\{ u^{a},\: n^{a}\right\} ,\left(u_{a}u^{a}=-n_{a}n^{a}=1,\quad u_{a}n^{a}=0\right),\label{1.3}\end{equation}
here $u^{a}$ is a tangent vector of the shell's world line so $u_{\pm}^{a}|_{_{\Sigma}}=dx^{a}/^{(2)}ds$,
and $n^{a}$ is the normal vector to the $\Sigma$ which is directed
from the region $D_{-}$ to the $D_{+}$. From the formula \eqref{1.3}
we find the equalities \begin{gather}
n_{0}=\sqrt{-g}u^{1},\qquad n_{1}=-\sqrt{-g}u^{0},\label{1.4}\\
u_{0}=\sqrt{-g}n^{1},\qquad u_{1}=-\sqrt{-g}n^{0}.\label{1.5}\end{gather}
We used $\gamma=\det\left\vert \gamma_{ab}\right\vert $.

\section{Equations of motion of a spherically-symmetric charged dust shell}

In the vacuum, the spherically-symmetric gravitational field of the
charged source is described by the Reissner-Nordstr\"{o}m metrics.
In curvature coordinates, we can write the metrics for the regions
$D_{-}$ and $D_{+}$ as follows\begin{equation}
^{(4)}ds_{\pm}^{2}=F_{\pm}dt_{\pm}^{2}-F_{\pm}^{-1}dr^{2}-r^{2}\left(d\Theta^{2}-\sin^{2}\Theta d\alpha^{2}\right),\label{2.1}\end{equation}
where \begin{equation}
F_{+}=1-\frac{2kM_{+}}{r}+\frac{kQ_{+}^{2}}{r^{2}},\qquad F_{-}=1-\frac{2kM_{-}}{r}+\frac{kQ_{-}^{2}}{r^{2}}.\label{2.2}\end{equation}
Here $t_{+}$ and $t_{-}$ are the Keeling time coordinates in the
exterior $D_{+}$ and interior $D_{-}$ regions, accordingly; $M_{+}$
and $M_{-}$ are the active masses; $Q_{+}$ and $Q_{-}$ are the
electric charges. These charges also generate electric fields with
potentials $\varphi_{\pm}=Q_{\pm}/r$ in regions $D_{\pm}$.

For the spherically-symmetric charged dust shell the motion equations
have the form (e.g. see \cite{deLaCruz})

\begin{gather}
n_{a}\frac{Du^{a}}{ds}|_{+}+n_{a}\frac{Du^{a}}{ds}|_{-}=\frac{2}{\sigma}\left[T_{\alpha\beta}n^{\alpha}n^{\beta}\right],\label{2.3}\\
n_{a}\frac{Du^{a}}{ds}|_{+}-n_{a}\frac{Du^{a}}{ds}|_{-}=4\pi\kappa\sigma=\frac{\kappa m}{r^{2}},\label{2.4}\end{gather}
where $T_{\alpha\beta}$ is the energy-momentum tensor, $m=4\pi\sigma r^{2}$
is the rest mass of the shell and $Du^{a}=u_{;b}^{a}dx^{b}$ is the
covariant differential relative to the metrics $\gamma_{ab}$. The
symbol $\left[\Phi\right]=\Phi|_{+}-\Phi|_{-}$ denotes the jump of
the quantity $\Phi$ on $\Sigma$. The signs {}``$|_{+}$'' and
{}``$|_{-}$'' indicate the marked quantities to be calculated as
the limit values when approaching the boundary $\Sigma$ from inside
and outside, respectively. In our case (see \cite{deLaCruz}): \begin{equation}
\left[T_{\alpha\beta}n^{\alpha}n^{\beta}\right]=\frac{q}{8\pi r^{4}}\left(Q_{+}+Q_{-}\right),\qquad q=Q_{+}-Q_{-},\label{2.5}\end{equation}
where $q=4\pi\sigma_e r^2$ is the charge of the shell.

Two-dimensional equations of motion \eqref{2.3} and \eqref{2.4}
allow us to obtain independent equations of motion of the shell in
the coordinates for each of the two-dimensional areas $D_{+}^{(2)}$
and $D_{-}^{(2)}$ separately. Taking into account the relations \eqref{1.4},
\eqref{1.5} and the equations \eqref{2.3}, \eqref{2.4} we obtain
the equations of motion of the shell in terms of quantities with respect
only to the $D_{+}^{(2)}$, or only to the $D_{-}^{(2)}$\begin{equation}
u_{;b}^{a}u^{b}|_{\pm}=\frac{Du^{a}}{ds}|_{\pm}=\frac{1}{m}\left(\frac{qQ_{\pm}}{r^{2}}\pm\frac{\kappa m^{2}-q^{2}}{2r^{2}}\right)n^{a}.\label{2.6}\end{equation}
We can see that regions $D_{+}^{(2)}$ and $D_{-}^{(2)}$, and their
boundaries $\Sigma_{\pm}^{(1)}$ together with the corresponding gravitational
and electric fields can be considered separately and independently,
as manifolds $D_{\pm}^{\left(2\right)}$ with edges $\Sigma_{\pm}^{(1)}$.

\section{The variational principle for a spherically-symmetric charged dust
shell}

Following \cite{gladVDNU}, we rewrite the equations of motion for
a shell \eqref{2.6} in the form, which is similar to the equations
of motion for a charge in an electromagnetic field\begin{equation}
u_{;b}^{a}u^{b}|_{\pm}=-G_{ab}^{\pm}u^{b}|_{\pm},\label{2.7}\end{equation}
where

\begin{equation}
G_{ab}^{\pm}=-G_{ba}^{\pm},\qquad G_{01}^{\pm}=\frac{1}{m}\left(-\frac{qQ_{\pm}}{r^{2}}\pm\frac{q^{2}-\kappa m^{2}}{2r^{2}}\right).\label{2.8}\end{equation}
To obtain this result the relation \eqref{2.5} has been used with
the fact that for the Reissner-Nordstr\"{o}m metrics we have $\sqrt{-\gamma}=1$.
The motion equations \eqref{2.7} define the trajectories $x_{\pm}^{a}=x_{\pm}^{a}(s)$
corresponding to real motions of the shell in the regions $D_{\pm}^{(2)}$.

Now we will consider other possible trajectories $\tilde{x}_{\pm}^{a}\left(s\right)=x_{\pm}^{a}\left(s\right)+\delta x_{\pm}^{a}\left(s\right)$,
which are sufficiently close to the real trajectory. From the relation
\eqref{2.7} it follows that\begin{equation}
\left(\left(u_{;b}^{a}u^{b}+G_{ab}u^{b}\right)\delta x^{a}\right)_{\pm}=0.\label{3.1}\end{equation}
This equation is a relativistic analogue of the D'Alambert principle
of virtual displacement when the time coordinate $x^{0}=$ $x^{0}\left(s\right)$
and the spatial one $x^{1}=$ $x^{1}\left(s\right)$ are considered
as the dynamic variables.

Let us multiply expressions \eqref{3.1} by $^{(2)}ds$ and integrate
the result along trajectories $\gamma_{\pm}$, then we get \begin{equation}
\intop_{\gamma_{\pm}}\left(\left(u_{;b}^{a}u^{b}+G_{ab}u^{b}\right)\delta x^{a}\right)_{\pm}^{(2)}ds=0.\label{3.2}\end{equation}
First term in this formula can be transformed using the variational
relation \begin{equation}
\delta|_{\pm}^{(2)}ds|_{\pm}=-\left(u_{;b}^{a}u^{b}\delta x^{a(2)}ds\right)|_{\pm}+d\left(u_{a}\delta x^{a}\right)|_{\pm}.\label{3.3}\end{equation}
Second term in the formula \eqref{3.2} can be written as follows
\begin{equation}
\left(G_{ab}u^{b}\delta x^{a}\right)_{\pm}^{(2)}ds=\left(G_{01}\left(dx^{0}\delta x^{1}-dx^{1}\delta x^{0}\right)\right)_{\pm}.\label{3.4}\end{equation}
In the regions $D_{\pm}^{(2)}$, let us introduce the auxiliary continuous
and invariant one-forms $\beta^{\pm}=B_{a}^{\pm}x^{a}$ by means of
the relations \begin{equation}
d\beta^{\pm}=G_{01}^{\pm}\left(dx^{0}\wedge dx^{1}\right)_{\pm}.\label{3.5}\end{equation}
Here $B_{a}^{\pm}=B_{a}^{\pm}\left(x^{0},x^{1}\right)$ are the vector
potentials of the gravitational and electric self-actions, and $G_{ab}^{\pm}\equiv B_{b,a}^{\pm}-B_{a,b}^{\pm}$
are it's intensities in the exterior and interior regions $D_{\pm}^{(2)}$,
respectively. Note that in two-dimensional space the integrability
condition for the relations \eqref{3.5} holds identically. Making
use of the definitions \eqref{3.5} and the formulae

\begin{equation}
d\left(B_{a}\delta x^{a}\right)_{\pm}-\delta\left(B_{a}\delta x^{a}\right)_{\pm}=G_{01}^{\pm}\left(dx^{0}\delta x^{1}-dx^{1}\delta x^{0}\right)_{\pm},\label{3.6}\end{equation}
we obtain second term in \eqref{3.2} as \begin{equation}
\left(G_{ab}u^{b}\delta x^{a}\right)|_{\pm}^{(2)}ds=\left(d\left(B_{a}\delta x^{a}\right)-\delta\left(B_{a}dx^{a}\right)\right)|_{\pm}.\label{3.7}\end{equation}
Substituting the expressions \eqref{3.3} and \eqref{3.7} into equation
\eqref{3.2}, we find the following variational formulae \begin{equation}
\delta\intop_{\gamma_{\pm}}\left(^{(2)}ds-B_{a}dx^{a}\right)_{\pm}+\left\{ \left(u_{a}+B_{a}\right)\delta x^{a}\right\} _{\pm}|_{A}^{B}=0,\label{3.8}\end{equation}
where indices $A$ and $B$ indicate that corresponding quantities
are taken in the initial and final position of the shell. Hence, it
follows that for all real trajectories $x_{\pm}^{a}=x_{\pm}^{a}(s)$
the integrals

\begin{equation}
I_{sh}^{\pm}=-m\intop_{\gamma_{\pm}}\left(^{(2)}ds-B_{a}dx^{a}\right)_{\pm}=-m\intop_{\gamma_{\pm}}\left(^{(2)}ds-\beta\right)_{\pm},\label{3.9}\end{equation}
have stationary values $\left(\delta I_{sh}^{\pm}=0\right)$ with
respect to arbitrary possible variations of the shell motions, when
initial and final positions remain fixed, i.e. $\left(\delta x_{\pm}^{a}\right)|_{A}^{B}=0$.
Thus, the requirement of stationarity $\delta I_{sh}^{+}=0$ and $\delta I_{sh}^{-}=0$
with respect to the arbitrary variations of the coordinates $\delta x^{+}$
and $\delta x^{-}$ yield to the equations of motion of a charged
dust shell \eqref{2.7} in the coordinates $x^{+}$ and $x^{-}$,
respectively. Hence, it can be seen that $I_{sh}^{-}$ is the effective
action for a charged dust shell defined only in the coordinates $x_{-}^{a}$
(i.e. in the interior region), and $I_{sh}^{+}$ is the effective
action for a charged dust shell defined only in coordinates $x_{+}^{a}$
in the exterior region.

The proposed procedure for the construction of the effective action
is the relativistic analogue of the classic method of deriving the
integral Hamilton principle from the D'Alembert principle of virtual
displacements \cite{Dalamb}.

\section{The effective actions and Lagrangians for a spherical charged dust
shell in curvature coordinates}

The effective actions \eqref{3.9} for a spherical charged dust shell
are obtained in the invariant form. Using curvature coordinates, we
choose common spatial spherical coordinates $\left\{ r,\theta,\alpha\right\} $
in $D_{\pm}$, and individual time coordinates $t_{\pm}$ in $D_{\pm}$,
respectively. Then the world sheet of the shell $\Sigma$ is given
by equations $r=r_{-}\left(t_{-}\right)$ and $r=r_{+}\left(t_{+}\right)$,
respectively.

In this case, from definitions \eqref{3.5} we obtain

\begin{equation}
d\beta^{\pm}=\left(G_{01}^{\pm}\left(r\right)dt\wedge dr\right)_{\pm}=-dt_{\pm}\wedge dV^{\pm}=d\wedge\left(V^{\pm}dt_{\pm}\right),\label{4.1}\end{equation}
where \begin{equation}
mV^{\pm}=-q\varphi_{\pm}\pm U,\label{4.2}\end{equation}
and

\begin{equation}
U=\frac{q^{2}-km^{2}}{2r}\label{4.5}\end{equation}
is the full effective potential energy of the gravitational and electromagnetic
self-actions of the shell. First term in \eqref{4.2} is the interaction
energy of the shell's charge $q$ and electric fields with potential
$\varphi_{\pm}=Q_{\pm}/r$ in regions $D_{\pm}^{(2)}$, respectively.
The general solutions of the equations in exterior derivatives \eqref{4.1}
for each of two regions $D_{\pm}^{(2)}$ can be written in the forms
\begin{equation}
\beta^{\pm}=V^{\pm}dt_{\pm}+d\psi_{\pm}=\frac{1}{m}\left(-q\varphi_{\pm}\pm U\right)dt_{\pm}+d\psi_{\pm},\label{4.3}\end{equation}
where $\psi_{\pm}=$ $\psi_{\pm}\left(t_{\pm},r\right)$ is a function
which sets calibration of the vector potential $B_{a}$ in the regions
$D_{\pm}^{(2)}$. These functions can be chosen so that one-form $\beta$
will be continuous on the shell (namely $\beta^{+}|_{_{\Sigma}}=\beta^{-}|_{_{\Sigma}}=\beta$).

Substituting one-form \eqref{4.3} into the actions \eqref{3.9},
we get the general representation of the effective actions for the
charged dust spherical shell in the form

\begin{equation}
I_{sh}^{\pm}=-\intop_{\gamma}\left\{ m^{(2)}ds+\left(q\varphi_{\pm}\mp U\right)dt-md\psi\right\} |_{\pm}.\label{4.4}\end{equation}
Making use of the gauge conditions $\psi_{\pm}=0$ in each of the
regions $D_{\pm}^{(2)}$ the effective actions for the charged dust
spherical shell can be written as \begin{equation}
I_{sh}^{\pm}=\intop_{\gamma}L_{sh}^{\pm}dt|_{\pm}=I_{sh}^{\pm}=-\intop_{\gamma}\left\{ m^{(2)}ds+\left(q\varphi_{\pm}\mp U\right)dt\right\} |_{\pm}.\label{4.6}\end{equation}
Here the effective Lagrangians have been introduced in the form \begin{equation}
L_{sh}^{\pm}=-m\left(\frac{^{(2)}ds}{dt}\right)_{\pm}-q\varphi_{\pm}\pm U=-m\sqrt{F_{\pm}-F_{\pm}^{-1}r_{t\pm}^{2}}-q\varphi_{\pm}\pm U.\label{4.7}\end{equation}
They describe dynamics of a charged dust spherical shell from the
point of view of the interior or exterior stationary observers. Here,
for the sake of simplification, the radial velocity is denoted by
$r_{t\pm}=dr/dt_{\pm}$. Note that one-form $\beta^{\pm}=\varphi^{\pm}\left(t_{\pm},r\right)dt_{\pm}$
is not continuous on the shell $\Sigma$ any more.

In the limiting case of small $m$ and $q$, it can be formally put
$M_{+}=M_{-}=M$, $Q_{+}=Q_{-}=Q$ and $U=0$. Then the Lagrangians
\eqref{4.7} will describe the test charged shell with mass $m$ and
charge $q$, which moves in the gravitational Reissner-Nordstr\"{o}m
field with parameters $M$ and $Q$, and in the electric field with
potential $\varphi=Q/r$.

\section{The isometry condition and the Hamiltonian constraint}

The effective actions $I_{sh}^{\pm}$ independently determine dynamics
of the shell in the regions $D_{\pm}^{(2)}$. Therefore, the regions
$D_{\pm}^{(2)}$ together with the boundaries $\Sigma_{\pm}^{\left(1\right)}$
and the corresponding fields can be considered separately and independently.
The boundaries $\Sigma_{\pm}^{\left(1\right)}$ acquire the physical
sense of the different faces of a dust shell with world sheet $\Sigma^{\left(1\right)}$
if regions $D_{\pm}^{(2)}$ are joined along these boundaries. However,
this requirement can be realized only if the condition of the isometry
for the boundaries $\Sigma_{\pm}^{\left(1\right)}$\begin{equation}
F_{+}dt_{+}^{2}-F_{+}^{-1}dr^{2}=F_{-}dt_{-}^{2}-F_{-}^{-1}dr^{2}=d\tau^{2},\label{6.1}\end{equation}
is fulfilled. Here $\tau$ is the proper time of the shell. Herewith,
$\Sigma_{+}^{(1)}=\Sigma_{-}^{(1)}=\Sigma^{(1)}$, $\gamma_{+}\left(t_{+}\right)=\gamma_{-}\left(t_{-}\right)=\gamma$.

It can be shown easily that the condition of isometry of the boundaries
\eqref{6.1} leads to the Hamiltonian constraints. First of all, we
have the relationships for the velocities\begin{gather}
\frac{F_{+}}{r_{t+}}-\frac{1}{F_{+}}=\frac{F_{-}}{r_{t-}}-\frac{1}{F_{-}},\label{6.2}\\
r_{\tau}^{2}\equiv\left(\frac{dr}{d\tau}\right)^{2}=\frac{r_{t\pm}^{2}}{F_{\pm}-F_{\pm}^{-1}r_{t\pm}^{2}},\qquad r_{t\pm}^{2}\equiv\left(\frac{dr}{dt_{\pm}}\right)^{2}=\frac{F_{\pm}^{2}r_{\tau}^{2}}{F_{\pm}+r_{\tau}^{2}}.\label{6.3}\end{gather}
Further, from the Lagrangians \eqref{4.7} we find the momenta and
Hamiltonians for the shell \begin{gather}
P_{\pm}=\frac{\partial L_{sh}^{\pm}}{\partial r_{t\pm}}=\frac{mr_{t\pm}}{F_{\pm}\sqrt{F_{\pm}-F_{\pm}^{-1}r_{t\pm}^{2}}}=\frac{m}{F_{\pm}}r_{\tau},\label{6.4}\\
H_{sh}^{\pm}=\sqrt{F_{\pm}\left(m^{2}+F_{\pm}P_{\pm}^{2}\right)}+q\varphi_{\pm}\mp U=E_{\pm}.\label{6.5}\end{gather}
Here $E_{\pm}$ are the energies, which are conjugated to the coordinate
times $t_{\pm}$ and are conserved in the frames of reference of the
respective stationary observers (interior or exterior one). Eliminating
the velocity $r_{\tau}$ from \eqref{6.4} and \eqref{6.5} , the
condition of isometry for the boundaries $\Sigma_{\pm}^{\left(1\right)}$can
be written as\begin{gather}
F_{+}P_{+}=F_{-}P_{-},\label{6.6}\\
\left(E_{+}-q\varphi_{+}+U\right)^{2}-m^{2}F_{+}=\left(E_{-}-q\varphi_{-}-U\right)^{2}-m^{2}F_{-}.\label{6.7}\end{gather}
Substituting $\varphi_{\pm}=Q_{\pm}/r$ into the last equation and
making use of the equations \eqref{2.2}, \eqref{4.5} we obtain \begin{equation}
H_{sh}^{+}=H_{sh}^{-}=M_{+}-M_{-}=E.\label{6.9}\end{equation}
Here $E=E_{+}=E_{-}$ denotes the total energy of the shell, which
is conjugated both to the coordinate times $t_{+}$ and $t_{-}$,
and its value is independent of the stationary observerпїЅs position
(inside or outside of the shell). It can be shown that the momentum
constraint \eqref{6.6} is a consequence of the relation \eqref{6.7}.

Thus, the dynamic system described by the Lagrangians $L_{sh}^{\pm}$
are not independent. They satisfy the Hamiltonian constraint \eqref{6.9},
which ensures the isometry of the shell faces.

Hamiltonian constraint \eqref{6.9} can be rewritten using the relation
\eqref{6.5} in terms of momenta in square form

\begin{equation}
F_{\pm}^{-1}\left(M_{+}-M_{-}-q\varphi_{\pm}\pm U\right)^{2}-F_{\pm}P_{\pm}^{2}=m^{2}.\label{6.10}\end{equation}
Hence, taking into account $P_{\pm}=-dS_{\pm}/dr$ we find the stationary
Hamilton-Jacobi equation

\begin{equation}
F_{\pm}^{-1}\left(M_{+}-M_{-}-q\varphi_{\pm}\pm U\right)^{2}-F_{\pm}\left(\frac{dS_{\pm}}{dr}\right)^{2}=m^{2},\label{6.10a}\end{equation}
where $S$ is the reduced action.

Now, we derive the first-order differential equations of motion for
the charged dust shells. For this purpose we rewrite the Hamiltonian
constraint \eqref{6.9} using the formulae \eqref{6.5} and \eqref{4.5}
in the form

\begin{equation}
m\sqrt{F_{\pm}+r_{\tau}^{2}}=\left[M\right]-\frac{qQ_{\pm}}{r}\pm\frac{q^{2}-km^{2}}{2r},\label{6.11}\end{equation}
or in the mixed form we have

\begin{equation}
m\sqrt{F_{-}+r_{\tau}^{2}}+m\sqrt{F_{+}+r_{\tau}^{2}}=2\left(M_{+}-M_{-}\right)-\frac{q\left(Q_{-}+Q_{+}\right)}{r},\label{6.12}\end{equation}

\begin{equation}
m\sqrt{F_{-}+r_{\tau}^{2}}-m\sqrt{F_{+}+r_{\tau}^{2}}=\frac{\kappa m^{2}}{r}.\label{6.13}\end{equation}

Note that this formulas are reasonable outside the event horizon,
where the curvature coordinates are valid. Formally, we can use these
formulas under the horizon too, i.e. in $T^{-}$- and $T^{+}$-regions,
assuming $r$ to be the time coordinate. It turns out that in order
to use the simplicity and convenience of the curvature coordinates
and to conserve the information about the shells in the region $R^{-}$,
it is sufficient to introduce an additional discrete variable $\epsilon=\pm1$
and perform the replacement $^{(2)}ds_{\pm}\rightarrow\epsilon_{\pm}^{(2)}ds_{\pm}$
in the actions $I_{sh}^{\pm}$ \eqref{4.6} (for more details on the
neutral shells, see \cite{gladJMa,gladGRG}). Here, $\epsilon_{\pm}=1$
corresponds to the region $R^{+}$, and $\epsilon_{\pm}=-1$ -- to
the region $R^{-}$. Then, for the extended system, Hamiltonians \eqref{6.5}
take the form

\begin{equation}
H_{sh}^{\pm}=\epsilon_{\pm}\sqrt{F_{\pm}\left(m^{2}+F_{\pm}P_{\pm}^{2}\right)}+q\varphi_{\pm}\mp U.\label{6.14}\end{equation}

\section{Special cases of dust shells}

\subsection{Hollow and screening shells}

Since isometry of the sides of the shell and the Hamiltonian constraint,
the considerations of the shell in terms of interior or exterior frames
of reference are equivalent. Therefore, we can choose the coordinates
in which the equations of motion have the most simple and most convenient
form. For example, the equations of motion of charged shells are greatly
simplified when one of the regions of space-time, inside or outside
of the shell, is flat. Thus, for ``hollow'' shell the coordinates
of the interior region are convenient, and for the ``screening''
shell so the coordinates of the exterior region are.

In the first case we have a self-gravitating shell, for which $M_{-}=0$
and $Q_{-}=0$. Such a shell, in the interior coordinates of flat
space-time, moves only under the influence of the potential energy
$U$ of the gravitational and the electric self-interactions \eqref{4.5},
which depends only on the rest mass $m$ and charge $q$ of the shell.
Let us use the following notations $M_{+}=M$ and $Q_{+}=Q$. In this
case, the exterior region $D_{+}$ of the shell is described by Reissner-Nordstr\"{o}m
metrics \eqref{2.1}, where $F_{+}=F=1-2kM/r+kQ^{2}/r^{2}$. In terms
of the coordinates $\left\{ t_{+,\,}r\right\} $, the Lagrangian,
the Hamiltonian and the Hamiltonian constraint can be written as

\begin{gather}
L_{sh}^{+}=-m\sqrt{F-F^{-1}r_{t_{+}}^{2}}-\frac{qQ}{r}+\frac{q^{2}-km^{2}}{2r}\,,\label{7.1}\\
H_{sh}^{+}=\sqrt{F\left(m^{2}+FP_{+}^{2}\right)}+\frac{qQ}{r}-\frac{q^{2}-km^{2}}{2r}=M\,,\label{7.1a}\\
P_{+}=\frac{mR_{t+}}{F\sqrt{F-F^{-1}R_{t+}^{2}}}\,,\label{7.2}\\
F^{-1}\left(M-q\varphi_{+}+\frac{q^{2}-km^{2}}{2r}\right)^{2}-FP_{+}^{2}=m^{2}\:.\label{7.3}
\end{gather}
In the interior region $D_{-}$ we have $F_{-}=1$. In terms of the
coordinates $\left\{ t_{-,\,}r\right\} $, the Lagrangian, the Hamiltonian
and the Hamiltonian constraint are much simpler and they have the
form

\begin{gather}
L_{sh}^{-}=-m\sqrt{1-r_{t-}^{2}}-\frac{q^{2}-km^{2}}{2r},\label{7.4}\\
H_{sh}^{-}=\sqrt{m^{2}+P_{-}^{2}}+\frac{q^{2}-km^{2}}{2r}=M\,,\qquad P_{-}=\frac{mr_{t-}}{\sqrt{1-r_{t-}^{2}}}\,,\label{7.5}\end{gather}
\begin{equation}
\left(M-\frac{q^{2}-km^{2}}{2r}\right)^{2}-P_{-}^{2}=m^{2}\:.\label{7.6}\end{equation}
In the case of the ``screening'' shell $M_{+}=0$, $Q_{+}=0$,
and we put $M_{-}=-M$, $Q_{-}=Q$. Thus, the system has a nontrivial
electric and gravitational fields only in the interior region $D_{-}$
of the shell. The easiest way is to describe such a shell in terms
of the coordinates $\left\{ t_{+,\,}r\right\} $ of the exterior region,
where it is moving under the influence of the potential energy of
the gravitational and the electric self-interaction of the same form
as for a {}``hollow'' shell, but with the opposite sign. Thus, we
have the Lagrangian, Hamiltonian and Hamiltonian constraint in the
form

\begin{equation}
L_{sh}^{+}=-m\sqrt{1-r_{t+}^{2}}+\frac{q^{2}-km^{2}}{2r}\,,\label{7.7}\end{equation}

\begin{equation}
H_{sh}^{+}=\sqrt{m^{2}+P_{+}^{2}}-\frac{q^{2}-km^{2}}{2r}=M\,,\qquad P_{+}=\frac{mr_{t+}}{\sqrt{1-r_{t+}^{2}}}\,,\label{7.8}\end{equation}

\begin{equation}
\left(M+\frac{q^{2}-km^{2}}{2r}\right)^{2}-P_{+}^{2}=m^{2}\,,\label{7.9}\end{equation}
correspondingly.

\subsection{A family of concentric charged dust shells}

Let us briefly consider a more complex configuration, consisting of
a set of concentric charged dust shells. Let $R_{a},\, m_{a},\, q_{a}\,,\tau_{a}$
be the radius, the proper mass, the charge, and the proper time of
the $a$-th shell, respectively $\left(a=1,2,...,N\right)$. We assume
that $R_{b}>R_{a}$ if $b>a$. Suppose that $M_{a},\, Q_{a}$ are
the active mass and the electric charge that determine the gravitational
Reissner-Nordstr\"{o}m field $F_{a}=1-2kM_{a}/r+kQ_{a}^{2}/r^{2}$
in the area $R_{a}<r<R_{a+1}$, between $a$-th and $(a+1)$-th shells.
We denote by $F_{a}^{-},\:\varphi_{a}^{-}$ and $F_{a}^{+},\:\varphi_{a}^{+}$
the metric coefficients and the electric potentials in the neighborhood
of $a$-th shell, in its interior $R_{a-1}<r<R_{a}$ and exterior
$R_{a}<r<R_{a+1}$ regions, respectively. Then\begin{gather}
F_{a}^{-}=1-\frac{2kM_{a-1}}{r}+\frac{kQ_{a-1}^{2}}{r^{2}},\qquad F_{a}^{+}=F_{a}=1-\frac{2kM_{a}}{r}+\frac{kQ_{a}^{2}}{r^{2}}\,,\label{7.10}\end{gather}
\begin{equation}
\varphi_{a}^{-}=\frac{Q_{a-1}}{r}\,,\qquad\varphi_{a}^{+}=\frac{Q_{a}}{r}\,,\qquad U_{a}=\frac{q_{a}^{2}-km_{a}^{2}}{2r}\,,\label{7.11}\end{equation}
where $U_{a}$ is the potential energy of self-interaction of the
$a$-th shell. Note that $F_{a}^{+}=F_{a+1}^{-}$, $\varphi_{a}^{+}=\varphi_{a+1}^{-}$
and $q_{a}=Q_{a}-Q_{a-1}$. In this case

\begin{equation}
H_{a}^{\pm}=\epsilon_{a}^{\pm}\sqrt{F_{a}^{\pm}\left(m^{2}+F_{a}^{\pm}\left(P_{a}^{\pm}\right)^{2}\right)}+q\varphi_{a}^{\pm}\mp U_{a}\label{7.12}\end{equation}
are the Hamiltonians of the $a$-th shell, which, as well as the momenta
of the shells\begin{equation}
P_{a}^{\pm}=\frac{m_{a}dr_{a}}{F_{a}^{\pm}d\tau_{a}}\,,\label{7.13}\end{equation}
are considered relatively to the coordinates of areas $R_{a-1}<r<R_{a}$
and $R_{a}<r<R_{a+1}$, respectively. Here $\epsilon_{a}^{\pm}=\pm1$.
These Hamiltonians satisfy the constraints\begin{equation}
H_{a}^{+}=H_{a}^{-}=M_{a}-M_{a-1}.\label{7.14}\end{equation}
The total Hamiltonian of the configuration\begin{equation}
H=\sum_{a=1}^{N}H_{a}^{\pm}=\sum_{a=1}^{N}\left\{ \epsilon_{a}^{\pm}\sqrt{F_{a}^{\pm}\left(m^{2}+F_{a}^{\pm}\left(P_{a}^{\pm}\right)^{2}\right)}+q\varphi_{a}^{\pm}\mp U_{a}\right\} ,\label{7.15}\end{equation}
by virtue of the Hamiltonian constraints \eqref{7.14} is provided
to be equal

\begin{equation}
H=E_{tot}=M_{N}-M_{0}.\label{7.15a}\end{equation}
Full electric charge of the configuration, because of the additivity
of the charge, is

\begin{equation}
Q=\sum_{a=1}^{N}q_{a}=Q_{N}-Q_{0}\end{equation}

If $M_{0}=0$ and $Q_{0}$, the system is moving in its own gravitational
and electric fields. In this case $H_{1}^{\pm}=M_{1}$, $Q_{1}=q_{1}$.
Thus, the full Hamiltonian and the charge of the system have the form\begin{equation}
H=M_{N}=M,\qquad Q=Q_{N},\label{7.16}\end{equation}
where $M=M_{N}$ is the full active mass of the configuration.

\section{Conclusions}

A special feature of the dynamics of the spherical shell is that its
evolution is not accompanied by radiation and can be reduced to a
simple Lagrangian system. This dynamic system has only one local degree
of freedom $r=r\left(\tau\right)$. Therefore, there is a possibility
to construct equations of motion of the shell in terms of coordinates
assigned only to the interior or exterior region and to consider them
independently. Hence, making use of the simple generalization of the
relativistic version of the D'Alambert principle of virtual displacements,
the effective actions $I_{sh}^{\pm}$ \eqref{4.6} are constructed
for a charged dust shell, describing its dynamics from the point of
view of the exterior or interior stationary observers. This leads
to the different effective Lagrangians $L_{sh}^{\pm}$ \eqref{4.7}
and Hamiltonians $H_{sh}^{\pm}$ \eqref{6.5} of
the shell in the interior and exterior regions $D_{\pm}^{\left(2\right)}$
with coordinates $x_{\pm}^{a}$. It turns out that the dynamical systems
described by these Lagrangians are not independent. They satisfy the
Hamiltonian constraint $H_{sh}^{+}=H_{sh}^{-}=M_{+}-M_{-}=E$, which
guarantees isometry of the sides of the shell. The total energy of
the shell $E=M_{+}-M_{-}$ conjugates both time coordinates $t_{\pm}$
in the regions $D_{\pm}$. Energy value is constant and it does not
depend on the position of a resting observer inside or outside of
the shell.

Consideration of the ``hollow'' and the ``screening'' charged
dust shells shows that their dynamics are somewhat similar. Also,
the generalized Hamiltonian constraint takes place for a family of
concentric spherical charged shells. Full Hamiltonian \eqref{7.15}
of the configuration numerically equals to the difference between
active masses outside the system and inside it, i.e. $H=E_{tot}=M_{N}-M_{0}$.

\section*{Acknowledgments}

This work was supported by the grant of the {}``Cosmomicrophysics''
programme of the Physics and Astronomy Division of the National Academy
of Sciences of Ukraine.

%\section*{References}


\begin{thebibliography}{99}

\bibitem{ansoldi02}
Ansoldi S 2002 WKB metastable quantum states of a
de Sitter-Reissner-Nordstr\"{o}m dust shell {\it Class. Quantum Grav.} {\bf 19} 6321-44

\bibitem{kijowski0}
Kijowski J, Magli G and Malafarina D 2009
The General solution for relativistic spherical shells
{\it Int.J.Mod.Phys. D} {\bf 18} 1801

\bibitem{israel} Israel W 1967 Singular hypersurfaces and thin shells
in general relativity {\it Nuovo Cimento} {\bf 44B} 1-14

\bibitem{kuchar} Kucha$\check{\mbox r}$ K 1968 Charged shells in general relativity
and their gravitational collapse  {\it Czech. J. Phys. B} {\bf 18} 435-63

\bibitem{deLaCruz} de la Cruz V and Israel W 1967 Gravitational Bounce
{\it Nuovo Cimento} {\bf 51A} 745-60

\bibitem{сhase} Chase J E 1970 Gravitational instability and collapse
of gharged fluid shells {\it Nuovo Cimento} {\bf 67B} 136-52

\bibitem{proszynski} Pr$\acute{\mbox o}$szy$\acute{\mbox n}$ski M  1978
Dynamics of a thin shell in Reissner-Nordstr\"{o}m geometry
{\it Acta Phys. Pol. B} {\bf 9} 613-616

\bibitem{farhi90} Farhi E, Guth A H and Guven J 1990
Is it possible to create a univrese in the labaratory by quantum tunnneling?
{\it Nuclear. Physics B} {\bf 339} 417-490

\bibitem{pre91}
Fischler W, Morgan D and Polchinski 1990
Quantization of false-vacuum bubbles: A Hamiltonian treatment of gravitational tunneling
{\it Phys. Rev. D} {\bf 42} 4042-4055
% Общий вар принцип и его редукция, время не конкретизируется!
% О паталогичности подхода квантования оболочки на фоне гравитации.

\bibitem{visser}
Visser M 1991 Quantum wormholes
{\it Phys. Rev. D} {\bf 43} 402-409
% Общий вар принцип и его редукция для заряженной оболочки, время Шварцшильдово внешнее!

\bibitem{berezin} Berezin V A, Boyarsky A M and Neronov A Yu 1998
Quantum geometrodynamics for black holes and wormholes
{\it Phys. Rev. D} {\bf 57} 1118-28
% Общий фор-м, действие по R, канон форм, Кучер переменные, канон перемен и констрэйт на %оболочке, квантование, уравнение Шредингера в конечных разностях

\bibitem{krauss} Kraus P and Wilczek F 1995 Self-interaction correction
to black hole radiance {\it  Nucl. Phys. B} {\bf 433} 403-20
% Общий фор-м, действие по R, лагранжиан, гамильтонов ф-м
%квантование, ВКБ, параболическое время

\bibitem{krauss2} Kraus P and Wilczek F 1995 Effect of self-interaction
on charged black hole radiance {\it Nucl. Phys. B} {\bf 437} 231-42
% Общий фор-м, действие по R, лагранжиан, гамильтонов ф-м
%квантование, ВКБ, параболическое время

\bibitem{ansoldi97} Ansoldi A, Aurilia A, Balbinot R and Spallucci E 1997
Classical and quantum shell dynamics and vacuum delay {\it Class. Quantum Grav.} {\bf 14} 2727-2756
% Общий вариационный принцип, полное действие из R, редуцированное действие, импульс %\tanh(), WKB solutions, WDW solution

\bibitem{alberghi99}
Alberghi G L, Casadio R, Vacca G P and Venturi G 1999
Gravitational collapse of a shell of quantized matter
{\it Class. Quantum Grav.} {\bf 16} 131-147
% Вариационный принцип, полное действие, редуцированное действие, импульс \tanh(), %собственное время, квантовая механика, WDW eq, WKB solutions

\bibitem{alberghi99a}
Alberghi G L, Casadio R and Venturi G 1999
Effective action and thermodynamics of radiating shells in general relativity
{\it Phys. Rev. D} {\bf 60} 124018-1-11

\bibitem{kijowski1}
H$\acute{\mbox a}$j$\acute{\mbox i}\check{\mbox c}$ek P and Kijowski J 1998
Lagrangian and Hamiltonian Formalism for Discontinuous Fluid and Gravitational Field
{\it Phys. Rev. D} {\bf 57} 914-935
%arXiv:gr-qc/9707020v1 9 Jul 1997

\bibitem{kijowski2} Kijowski J  1998
``True degrees of freedom'' of a spherically symmetric, self-gravitating dust shell
{\it Acta Phys. Pol. B} {\bf 29} 1001-1013

\bibitem{kijowski3}
H$\acute{\mbox a}$j$\acute{\mbox i}\check{\mbox c}$ek P, Kijowski J 2000
Spherically symmetric dust shell and the time problem in canonical relativity
{\it Phys. Rev. D} {\bf 62} 044025-1-044025-5
%(две квантовые механики основанные на этих двух временах, не могут, поэтому, быть unitarily эквивалентом.)
% the ‘‘royal road’’ to Hamiltonian reduction
%Another way is based on gauge fixing in the sense that we single out a representative
%in each equivalence class

\bibitem{kijowski4} Kijowski J and Czuchry E 2005
Dynamics of a self-gravitating shell of matter
{\it Phys. Rev. D} {\bf 72} 084015-1-084015-12
%e-Print: gr-qc/0507074

\bibitem{kijowski5}
Kijowski J, Magli G and Malafarina D 2006
New derivation of the variational principle for the dynamics of a gravitating spherical shell {\it Phys. Rev. D} {\bf 74} 084017-1-084017-11

\bibitem{gladJMa} Gladush V D 2001 On the variational principle
for dust shells in General Relativity {\it J. Math. Phys.} {\bf 42} 2590-610

\bibitem{gladGRG} Gladush V D 2004 The variational principle and
effective action for a spherical dust shel
{\it Gen. Rel. Grav.} {\bf 36} 1821-39

%\bibitem{carlip} Carlip S  Lectures in (2+1)-Dimensional Gravity 1995
%gr-qc/9503024

\bibitem{frolov98}
Frolov V P and Novikov I D 1998
{\it Black Hole Physics: Basic Concepts and New Developments}
Dordrecht, Netherlands: Kluwer Academic 770

\bibitem{frolov} Frolov V P (1974)
Motion of the charged radiative shells in General Relativity and the friedmon states
{\it Sov. Phys. JETP.} {\bf 39} 393
%Фролов~В.П. Движение заряженных излучающих оболочек в ОТО и фридмонные состояния
%% (ЖЭТФ, 66, 813-825, 1974).

\bibitem{berezin88}
Berezin V A, Kozimirov N G, Kuzmin V A, and Tkachev I I 1988
On the quantum mechanics of bubbles
{\it Physics Letters B} {\bf 212} 415-417


\bibitem{hajicek}
H$\acute{\mbox a}$j$\acute{\mbox i}\check{\mbox c}$ek P,
Kay B.S  and  Kucha$\check{\mbox r}$ K 1992
Quantum collapse of a self-gravitating: Equivalence to electric scattering
{\it  Phys. Rev. D} {\bf 46} 5439-48

\bibitem{nakamura}
Nakamura K, Oshiro Y and Tomimatsu A. 1996
Quantum formation of black hole and wormhole in gravitational collapse of a dust shell
{\it  Phys. Rev. D} {\bf 53} 4356-4365

\bibitem{dolgov} Dolgov A D and Khriplovich I B 1997
Instructive properties of quantized gravitating dust shell
{\it Phys. Lett. B} {\bf 400} 12-14

\bibitem{berezin97a}
Berezin V A 1997 Square-root quantization: application to quantum black holes
{\it Nucl. Phys. Proc. Suppl. D} {\bf 57} 181-183

\bibitem{berezin97}
Berezin V A 1997 Quantum black hole model and Hawking's radiation
{\it Phys. Rev. D} {\bf 55} 2139-2151

%\bibitem[15]{gladJMo} Gladush V D The quasi-classical model of
%the spherical configuration in general relativity 2002 {\it Int. J. Mod. Phys. D} {\bf 11} 367-89

\bibitem{Dalamb} D ter Haar 1971 {\it Elements of Hamiltonian mechanics} (Oxford: Pergamon)
\bibitem{Lanczos} Lanczos C 1972 {\it The variational principles of mechanics} (Toronto)

\bibitem{gladVDNU} Gladush V D, Martinenko V G and Rogoza B E 2006
Relativistic analogue of the D'Alembert principle of virtual displacement
and effective action for the dust spherical shell in General Relativity
{\it Vistnik Dniepropetrovsk University} {\bf 13} 91-97


\end{thebibliography}
\end{document}